\documentclass{iau} 
\usepackage{graphicx}
\usepackage{natbib}

\title[IAU324. Quasar X-ray spectral variability] %% give here short title %%
{Ensemble quasar spectral variability from the XMM-Newton Serendipitous Source Catalogue}

\author[R. Serafinelli, F. Vagnetti and R. Middei]   %% give here short author list %%
{Roberto Serafinelli$^{1}$, Fausto Vagnetti$^{1}$ \and Riccardo Middei$^{2}$}

\affiliation{$^{1}$Dipartimento di Fisica, Universit\`a di Roma ``Tor Vergata''\\ Via della Ricerca Scientifica 1, 00133, Rome, Italy \\[\affilskip] $^{2}$Dipartimento di Matematica e Fisica, Universit\`a Roma Tre \\Via della Vasca Navale 84, 00146, Rome, Italy}

\pubyear{2016}
\volume{324}  %% insert here IAU Symposium No.
\jname{New Frontiers on Black Hole Astrophysics}
\editors{A.C. Editor, B.D. Editor \& C.E. Editor, eds.}
\begin{document}

\maketitle

\begin{abstract}
Variations of the X-ray spectral slope have been found in many Active Galactic Nuclei (AGN) at moderate luminosities and redshifts, typically showing a ``softer when brighter'' behaviour. However, similar studies are not usually performed for high-luminosity AGNs. We present an analysis of the spectral variability based on a large sample of quasars in wide intervals of luminosity and redshift, measured at several different epochs, extracted from the fifth release of the XMM Newton Serendipitous Source Catalogue. Our analysis confirms a ``softer when brighter'' trend also for our sample, extending to high luminosity and redshift the general behaviour previously found. These results can be understood in light of current spectral models, such as intrinsic variations of the X-ray primary radiation, or superposition with a constant reflection component.
\end{abstract}

\firstsection % if your document starts with a section,
\section{Introduction}
The spectral slope optical/UV variations in AGNs have been quantified by \cite{trevese02}, by means of the \textit{spectral variability parameter} $\beta=\Delta\alpha / \Delta\log F$, $\alpha$ being the slope of the spectrum and $F$ its flux in the given band. A positive value was found, which means that the spectrum is harder when the flux is higher.\\
However, in the X-ray band, the opposite behaviour has been found for individual sources, and very few systematic studies have been performed, e.g. \cite{sobolewska09}, who found this trend for a sample of 10 nearby Seyfert galaxies.\\
In our study we investigated the spectral variability of quasars, using the MEXSAS catalogue \citep{vagnetti16}, created cross-matching the multi-epoch observations of the XMM-Newton Serendipitous Source Catalogue, Data Release 5 \citep{rosen16} with two partially overlapping Sloan Digital Sky Survey catalogues, SDSS-DR7Q \citep{schneider10} and SDSS-DR12Q \citep{paris16}, obtaining a catalogue of 7,837 X-ray observations of 2,700 quasar sources.

\section{Ensemble and single source analyses}
\begin{figure}
\centering
\includegraphics[scale=0.37]{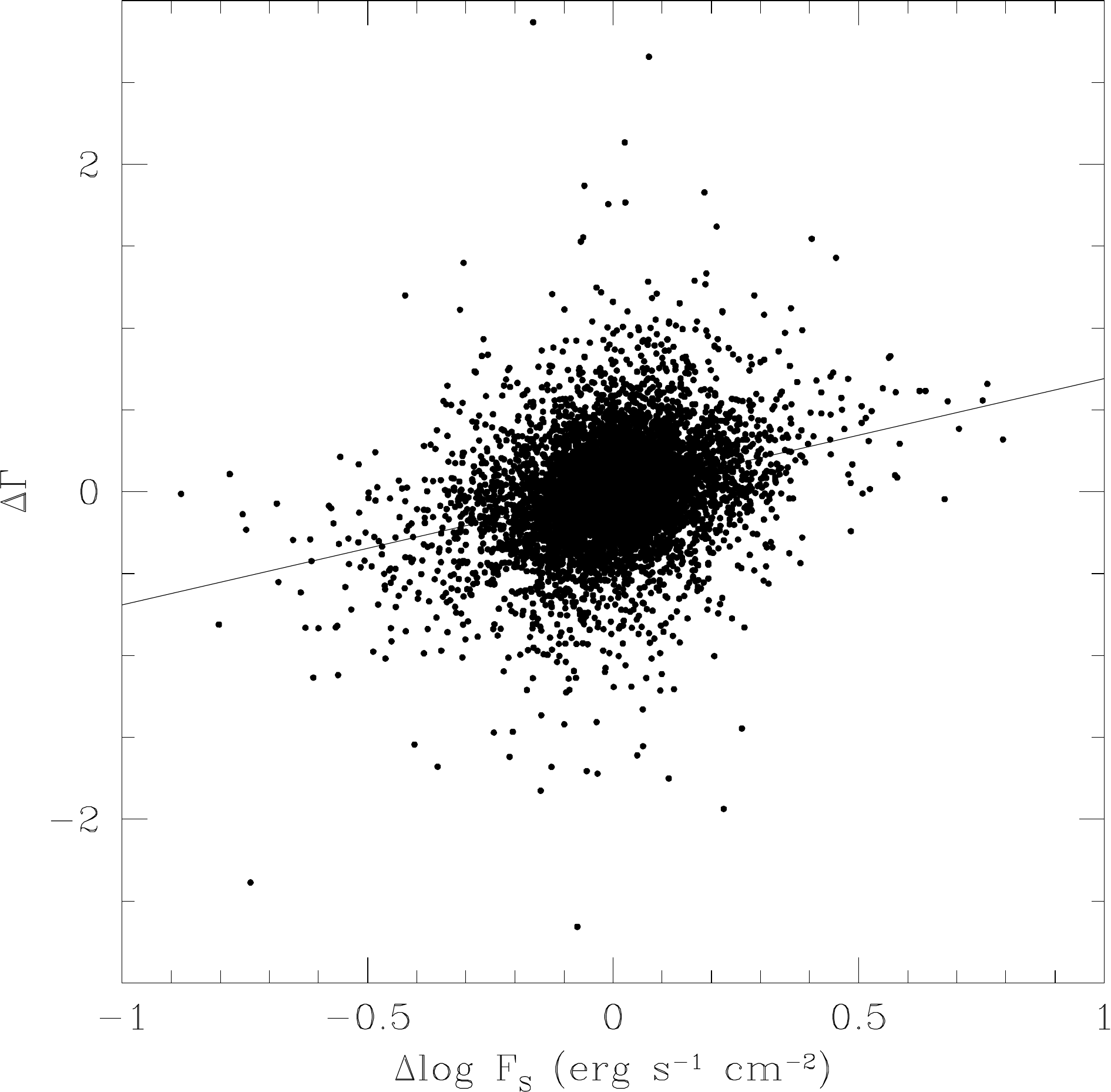}
\caption{$\Delta\Gamma-\Delta\log F_S$ (band $0.5-2$keV) correlation of the whole sample.}
\label{fig:ensemble}
\end{figure}

In order to study the spectral variability in the X-ray band, we redefine the spectral variability parameter in terms of the photon index $\Gamma$, defined after $N(E)\propto E^{-\Gamma}$. Then, the spectral variability parameter becomes $\beta=-\Delta\Gamma/\Delta\log F$, since $\Gamma=1-\alpha$. In order to better compare sources we computed the linear fit between the variations of $\Gamma$ and $\log F_S$ ($F_S$ in band $0.5-2$keV) from the source mean values. The computed spectral variability parameter is $\beta=-0.69\pm0.03$ (see Fig.~\ref{fig:ensemble}). A negative $\beta$ implies that the spectral slope becomes higher for increasing flux, making the spectrum steeper. This means that the spectrum is softer when brighter.\\
\indent We tried to investigate the dependence of $\beta$ from some source parameters, such as black hole mass, Eddington ratio, redshift and X-ray luminosity, finding no evidence of such dependence. The value of $\beta$ in some bins deviates significantly from the ensemble value, though, suggesting that different sources may have different values of $\beta$ between each other. Therefore we studied single sources, selecting the ones with low probability of finding a $\Gamma-\log F_S$ correlation by chance ($p\leq10^{-3}$). We find all negative $\beta$s in a wide range of values, from $\beta=-3.54\pm0.54$ to $\beta=-0.62\pm0.08$.

\section{Discussion}
The softer when brighter trend, found in both the ensemble and single-source analysis, may be caused by the superposition of the primary X-ray emission with an additional reflected component. According to some models \cite[e.g.,][]{shih02} the primary emission could be variable in flux, but not in spectrum, while the constant reflected component is not variable at all, producing the observed spectra. Other models \cite[e.g.,][]{liang79},  suggest that the primary component may be variable in spectrum as well.\\
As for the single source analysis, the wide range of $\beta$ values may be dependent on several unconsidered features of the source, such as black hole spin, angle of view or radio-loudness, or it may be influenced by stochastic processes.


\begin{thebibliography}{}

\bibitem[Liang (1979)]{liang79} Liang, E. P. T., 1979, \textit{ApJ}, \textbf{231}, L111.
\bibitem[P\^aris et al. (2016)]{paris16} P\^aris, I., Petitjean, P., Ross, N. P. et al., 2016, \texttt{arXiv:1608.06483}
\bibitem[Rosen et al. (2016)]{rosen16} Rosen, S. R., Webb, N. A., Watson, M. G. et al., 2016, \textit{A\&A}, \textbf{590}, A1.
\bibitem[Schneider et al. (2010)]{schneider10} Schneider, D. P., Richards, G. T., Hall, P. B., 2010, \textit{AJ}, \textbf{139}, 2360.
\bibitem[Shih et al (2002)]{shih02} Shih, D. C., Iwasawa, K., Fabian, A. C., 2002, \textit{MNRAS}, \textbf{333}, 687.
\bibitem[Sobolewsa and Papadakis (2009)]{sobolewska09} Sobolewska, M. A., Papadakis I. E., 2009, \textit{MNRAS}, \textbf{399}, 1597.
\bibitem[Trevese and Vagnetti (2002)]{trevese02} Trevese, D., Vagnetti, F., 2002, \textit{ApJ}, \textbf{564}, 624.
\bibitem[Vagnetti et al. (2016)]{vagnetti16} Vagnetti, F., Middei, R., Antonucci, M. et al, 2016, \textit{A\&A}, \textbf{593}, A55.


\end{thebibliography}
\end{document}